\documentclass[aps,prl,reprint,superscriptaddress]{revtex4-1}

\usepackage{graphicx}
\usepackage{color}
\usepackage{textcomp}
\usepackage{siunitx}
\usepackage{xspace}
 \usepackage{mathtools}

\DeclarePairedDelimiter\abs{\lvert}{\rvert}%

\newcommand{\ie}{i.\,e.\xspace}

\newcommand{\eg}{e.\,g.\xspace}

\newcommand{\Tn}{T$_0$\xspace}

\begin{document}


\title{Anisotropic Pauli spin blockade in hole quantum dots}



\author{Matthias Brauns}
\email[Corresponding author, e-mail: ]{m.brauns@utwente.nl}
\author{Joost Ridderbos}
\affiliation{NanoElectronics Group, MESA+ Institute for Nanotechnology, University of Twente, P.O. Box 217, 7500 AE Enschede, The
Netherlands}
\author{Ang Li}
\affiliation{Department of Applied Physics, Eindhoven University of Technology, Postbox 513, 5600 MB Eindhoven, The Netherlands}
\author{Erik P. A. M. Bakkers}
\affiliation{Department of Applied Physics, Eindhoven University of Technology, Postbox 513, 5600 MB Eindhoven, The Netherlands}
\affiliation{QuTech and Kavli Institute of Nanoscience, Delft University of Technology, 2600 GA Delft, The Netherlands}
\author{Wilfred G. van der Wiel}
\affiliation{NanoElectronics Group, MESA+ Institute for Nanotechnology, University of Twente, P.O. Box 217, 7500 AE Enschede, The
Netherlands}
\author{Floris A. Zwanenburg}
\affiliation{NanoElectronics Group, MESA+ Institute for Nanotechnology, University of Twente, P.O. Box 217, 7500 AE Enschede, The
Netherlands}


\date{\today}

\begin{abstract}
We present measurements on gate-defined double quantum dots in Ge-Si core-shell nanowires, which we tune to a regime with visible shell filling in both dots. We observe a Pauli spin blockade and can assign the measured leakage current at low magnetic fields to spin-flip cotunneling, for which we measure a strong anisotropy related to an anisotropic $g$~factor. At higher magnetic fields we see signatures for leakage current caused by spin-orbit coupling between (1,1)-singlet  and (2,0)-triplet states. Taking into account these anisotropic spin-flip mechanisms, we can choose the magnetic field direction with the longest spin lifetime for improved spin-orbit qubits.
\end{abstract}


\maketitle

\section{Introduction}

For quantum computation \citep{Aaronson2013quantum,DiVincenzo1995,Ladd2010}, increasing research efforts have focused in recent years on C, Si, and Ge \citep{Laird2014,Zwanenburg2013,Amato2014} because these materials can be purified to only consist of isotopes with zero nuclear spin \citep{Itoh2003, Itoh1993} and thus exhibit exceptionally long spin lifetimes \citep{Muhonen2014a,Veldhorst2014}. The one-dimensional character of Ge-Si core-shell nanowires leads to unique electronic properties in the valence band, where  heavy and light hole states are mixed \citep{Csontos2007,Csontos2009,Kloeffel2011a}. The band mixing gives rise to an enhanced Rashba-type spin-orbit interaction (SOI) \citep{Kloeffel2011a}, leading to strongly anisotropic and electric-field dependent $g$~factors \citep{Maier2013,Brauns2015}. This makes quantum dots in Ge-Si core-shell nanowires promising candidates for robust spin-orbit qubits. 
\paragraph{}
Crucial steps towards high-fidelity qubits are spin readout via a Pauli spin blockade (PSB) \citep{Ono2002} and understanding the spin relaxation mechanisms. For Ge-Si core-shell nanowires, there have been reports on charge sensing \citep{Hu2007} spin coherence \citep{Higginbotham2014a}, and spin relaxation \citep{Hu2012}. The authors of the latter performed their experiments along a single magnetic field axis and concluded additional work was needed to pinpoint the spin relaxation mechanism.
\paragraph{}
In this Rapid Communication, we define a hole double quantum dot in a Ge-Si core-shell nanowire by means of gates. We observe shell filling and a Pauli spin blockade. The measured leakage currents strongly depend on both the magnitude and direction of the magnetic field and are assigned to spin-flip cotunneling processes for low magnetic fields. We find signatures of SOI-induced leakage current at higher magnetic fields up to 1~T.

\section{Shell filling and Pauli spin blockade}

\begin{figure}
\includegraphics{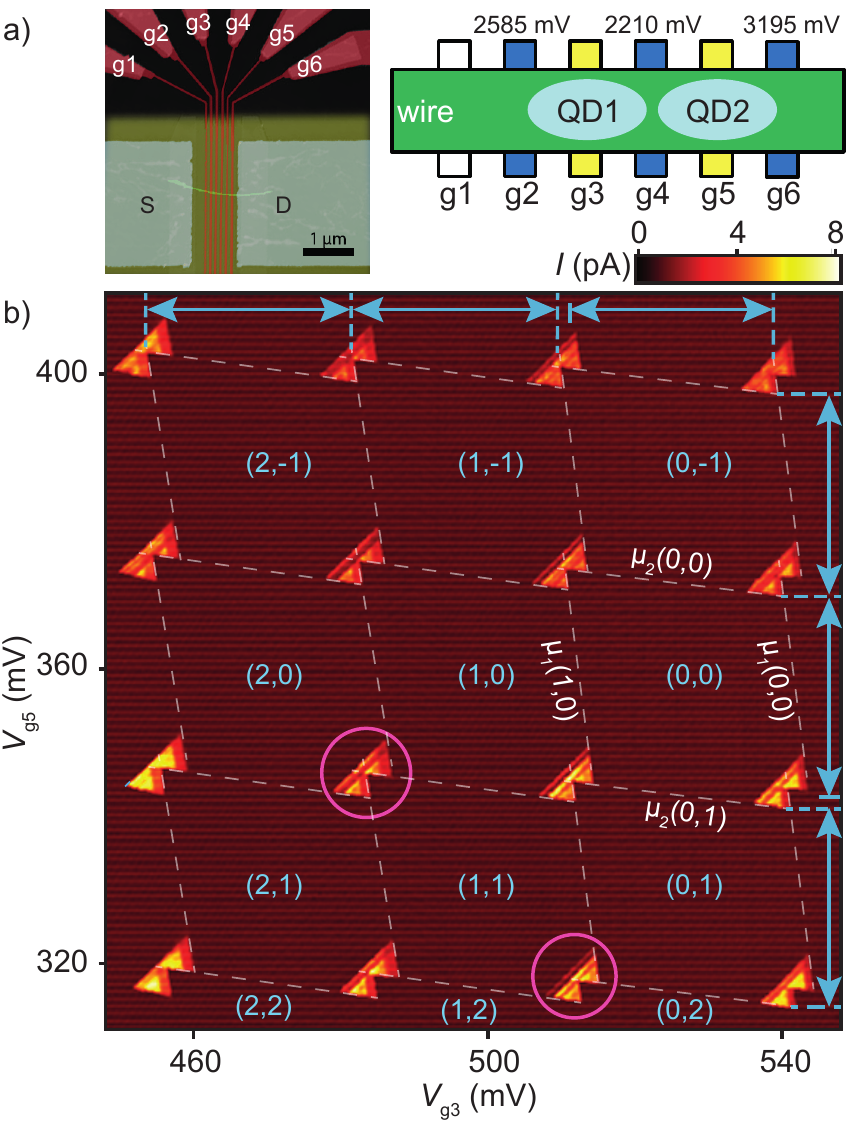}%
\caption{(a) False-color atomic-force microscopy image of the device (left) and schematic depiction of the gate configuration (right). (b) Current $I$ vs. $V_{\text{g3}}$ and $V_{\text{g5}}$ with $V_{\text{g2}} = 2585$~mV, $V_{\text{g4}} = 2210$~mV, $V_{\text{g6}} = 3195$~mV, and $V_{\text{SD}} = 2$~mV. White dashed lines are guides to the eyes for the honeycomb edges. Blue arrows represent $E_{\text{C}}$, and the gaps between adjacent arrows indicate an additional $E_{\text{orb}}$. Circles mark triangle pairs exhibiting PSB. $(m,n)$ denotes the effective hole occupation $m$ and $n$ on the left and right dot, respectively.\label{Fig1}}
\end{figure}

\begin{figure}
\includegraphics{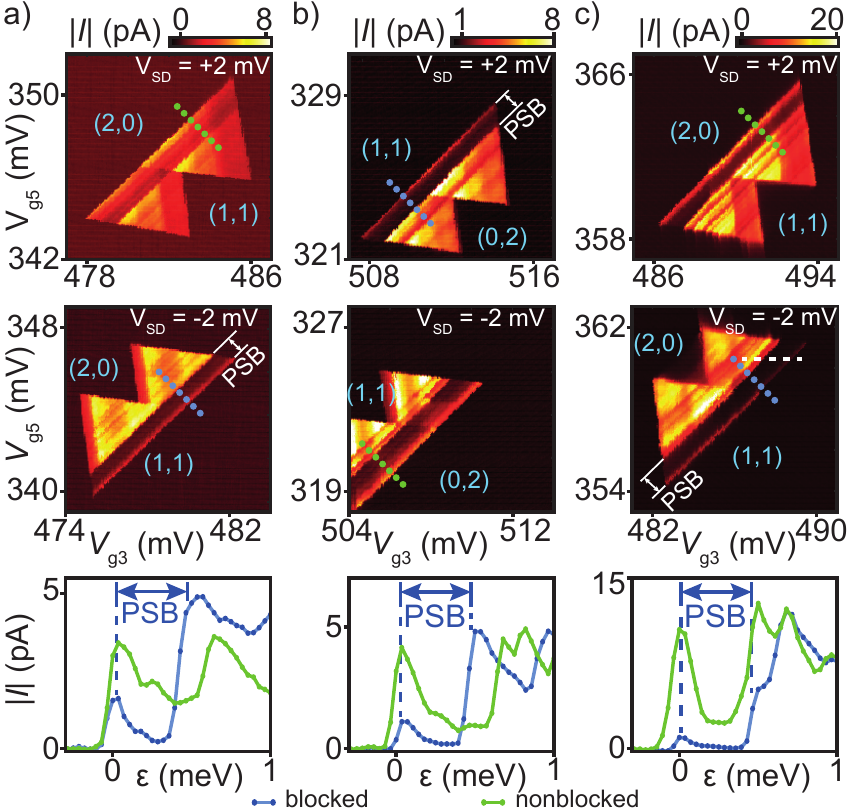}%
\caption{Zoom of the bias triangle pairs exhibiting PSB: (a) and (b) with the same barrier gate voltages as in Fig.~\ref{Fig1}, and (c) at $V_{\text{g2}} = 2570$~mV, $V_{\text{g4}} = 2210$~mV, and $V_{\text{g6}} = 3160$~mV. Line cuts (lowest panels) taken along the dotted lines for the $V_{\text{SD}}$ direction with (blue) and without PSB (green).\label{Fig2}}
\end{figure}

Our device in Fig.~\ref{Fig1}(a) consists of a \emph{p}$^{\text{++}}$-doped Si substrate covered with 200~nm SiO$_{\text{2}}$, on which six bottom gates with 100~nm pitch are patterned with electron-beam lithography (EBL). The gates are buried by 10~nm Al$_{\text{2}}$O$_{\text{3}}$ grown with atomic layer deposition at $100^{\circ}$C. A single nanowire with a Si shell thickness of $\sim$2.5~nm and a defect-free Ge core with a radius of $\sim$8~nm \citep{Li2016}, in preparation) is deterministically placed on top of the gate structure with a micromanipulator and then contacted with ohmic contacts made of 0.5/50~nm Ti/Pd. Based on transmission electron microscopy studies of similar nanowires the wire axis is most likely pointed along the $<$110$>$ crystal axis. A source-drain voltage $V_{\text{SD}}$ is applied between the source and ground, and the current $I$ is measured at the drain contact. All measurements are performed using direct current (dc) electronic equipment in a dilution refrigerator with a base temperature of 8~mK.
\paragraph{}
In Fig.~\ref{Fig1}(b) we plot $I$ versus the voltage $V_{\text{g3}}$ on gate g3 and the voltage $V_{\text{g5}}$ on g5. We see a highly regular pattern of bias triangles \cite{Wiel2002}, indicating the formation of a double quantum dot above gates g3 (``left dot'') and g5 (``right dot''). The 16 bias triangle pairs all have very sharp edges, and the absence of current along the honeycomb edges indicates a double quantum dot weakly coupled to the reservoirs \cite{Wiel2002}. We introduce $(m,n)$ as the effective charge occupation numbers $m$ and $n$ of the left and right dot, respectively.
\paragraph{}
Nine honeycomb cells are visible in Fig.~\ref{Fig1}(b). In each column from right to left a hole is added to the left dot, while in each row from top to bottom a hole is added to the right dot. The addition energy $E_{\text{add}}$ for each added hole can be extracted from Fig.~\ref{Fig1}(b) by measuring the distance between the triple points that are connected by the dashed lines and converting this distance graphically from the gate voltage into energy using the bias triangle size as a scale. The addition energy of the left dot is $E_{\text{add}} = 9.8\pm0.1$~meV for the middle and left column, and $E_{\text{add}} = 10.3\pm0.1$~meV for the right column. For the right dot $E_{\text{add}} = 9.5\pm0.1$~meV in the top and bottom row, but $E_{\text{add}} = 10.2\pm0.1$~meV in the middle row. The increased $E_{\text{add}}$ in the right column and middle row can be readily explained by the filling of a new orbital in the corresponding dot, so that in addition to the (classical) charging energy $E_{\text{C}}$ the (quantum-mechanical) orbital energy $E_{\text{orb}}$ has to be taken into account, with $E_{\text{orb}} = 0.5\pm0.1$~mV in the left and $E_{\text{orb}} = 0.7\pm0.1$~mV in the right dot. This filling of a new orbital for both dots allows us to identify the charge occupation $(m,n)$ with the occupation numbers of the newly filled orbitals in the left and right dot for $m,n\ge0$.
\paragraph{}
For spin-$\frac{1}{2}$ particles filling the orbitals, the bias triangle pairs marked by red circles in Fig.~\ref{Fig1}(b) should exhibit PSB in opposite $V_{\text{SD}}$ directions. Figures~\ref{Fig2}(a) and (b) display zooms of these triangle pairs for positive and negative $V_{\text{SD}}$. The lower panels of Fig.~\ref{Fig2}(a) and (b) display line cuts along the dotted lines for the $V_{\text{SD}}$ direction with (blue) and without (green) PSB. A suppression of the current at a detuning $\epsilon$ lower than the singlet-triplet splitting $\Delta_{\text{S-T}}$ can be observed at negative $V_{\text{SD}}$ in Fig.~\ref{Fig2}(a), where the zero-detuning current at positive $V_{\text{SD}}$ of $I_{\text{pos}}(\epsilon\!=\!0) = 3.6$~pA is reduced to $I_{\text{neg}}(\epsilon\!=\!0) = -1.6$~pA at negative $V_{\text{SD}}$. Current suppression at positive $V_{\text{SD}}$ is visible in Fig.~\ref{Fig2}(b), where $I_{\text{pos}}(\epsilon\!=\!0) = 1.1$~pA and $I_{\text{neg}}(\epsilon\!=\!0) = -4.2$~pA, exactly as expected. The current is thus only partially suppressed; comparable values for the resonant ($\epsilon\!=\!0$) leakage current in double quantum dots have been reported in the literature \citep{Pfund2007, Churchill2008, Nadj-Perge2010}. From the line traces taken in the spin-blocked bias direction, we extract a singlet-triplet splitting $\Delta_{\text{S-T}} = 0.5\pm0.1$~meV. Note that this value is very close to $E_{\text{orb}}$ (see above). This is reasonable since the (2,0) and (0,2) singlet-triplet splittings involve states originating from successive quantum dot orbitals.

\paragraph{}
We retune the device by lowering $V_{\text{g2}}$ and $V_{\text{g6}}$ in small, controlled steps and following the bias triangle pair at the (1,1)--(2,0) degeneracy in $V_{\text{g3}}$-$V_{\text{g5}}$ gate space [Fig.~\ref{Fig2}(c)]. From the line cuts (lower panel) we extract $I_{\text{pos}}(\epsilon\!=\!0) = 10.4$~pA and $I_{\text{neg}}(\epsilon\!=\!0) = 1.2$~pA, \ie, the non-blocked current is almost threefold higher after retuning, whereas the leakage current even reduced so that current rectification is now more pronounced.
\paragraph{}
In conclusion, Figs.~\ref{Fig1} and \ref{Fig2} display the formation of a gate-defined double quantum dot, in which we show orbital shell filling for both dots and extract orbital energies of 500--700~$\mu$eV. We find PSB for two bias triangle pairs in opposite bias directions.

\section{Anisotropic leakage current}

Our quantum dots are elongated along the nanowire axis $\vec{a}_\text{NW}$ and are exposed to a static electric field $\vec{E}$ from the bottom gates pointing out of the chip plane. We explore the origin of the leakage current $I_{\text{leak}}$ in PSB by performing magnetospectroscopy measurements along the white dashed line in Fig.~\ref{Fig2}(c) and plot $I_{\text{leak}}$ versus the detuning and the magnetic field in Fig.~\ref{Fig3}(a). In order to investigate possible anisotropic effects, we conduct measurements along three orthogonal directions of $\vec{B}$ [see also Fig.~\ref{Fig3}(d)]: ($B_z$) $\vec{B}$ parallel to the nanowire, ($B_y$) $\vec{B}$ perpendicular to both the nanowire and the electric field, and ($B_x$) $\vec{B}$ perpendicular to the nanowire and parallel to the electric field. 
\paragraph{}
In Fig.~\ref{Fig3}(a) we plot $I_{\text{leak}}$ vs $\epsilon_0$ and $B \equiv \abs{\vec{B}}$. Here, $\epsilon_0 \equiv \epsilon(B\!=\!0)$ is introduced as an absolute energy scale, since $\epsilon$ is only defined relative to the alignment of the (2,0) and (1,1) ground states. For all three $I_{\text{leak}}(\epsilon_0, B)$ plots in Fig.~\ref{Fig3}(a), we scan the $I_{\text{leak}}(\epsilon_0)$ line traces and set the center of the lowest-energy peak as $\epsilon\!=\!0$. In Figs.~\ref{Fig3}(b) and (c) we plot line traces for the leakage current at constant detuning as indicated in the panels. The green dashed lines in Fig.~\ref{Fig3}(a) are guides to the eye for $\epsilon(B) = 0$ (`S-onset'). The lifting of the blockade at $\epsilon=\Delta_{\text{S-T}}$ is indicated by the blue dashed lines (`\Tn-onset'). The shift of the S-onset and the \Tn-onset to positive $\epsilon_0$-values for increasing $B$ in the plots has also been observed in other experiments \citep{Kyriakidis2001,Fujisawa2003,Johnson2005} and is explained by orbital effects \citep{Wagner1992,VanderWiel1998,Kyriakidis2001}. A more detailed discussion can be found in the Supplemental Material S1 \citep{SuppMat}.

\subsection{Magnetospectroscopy along $B_z$}
\begin{figure}
	\includegraphics{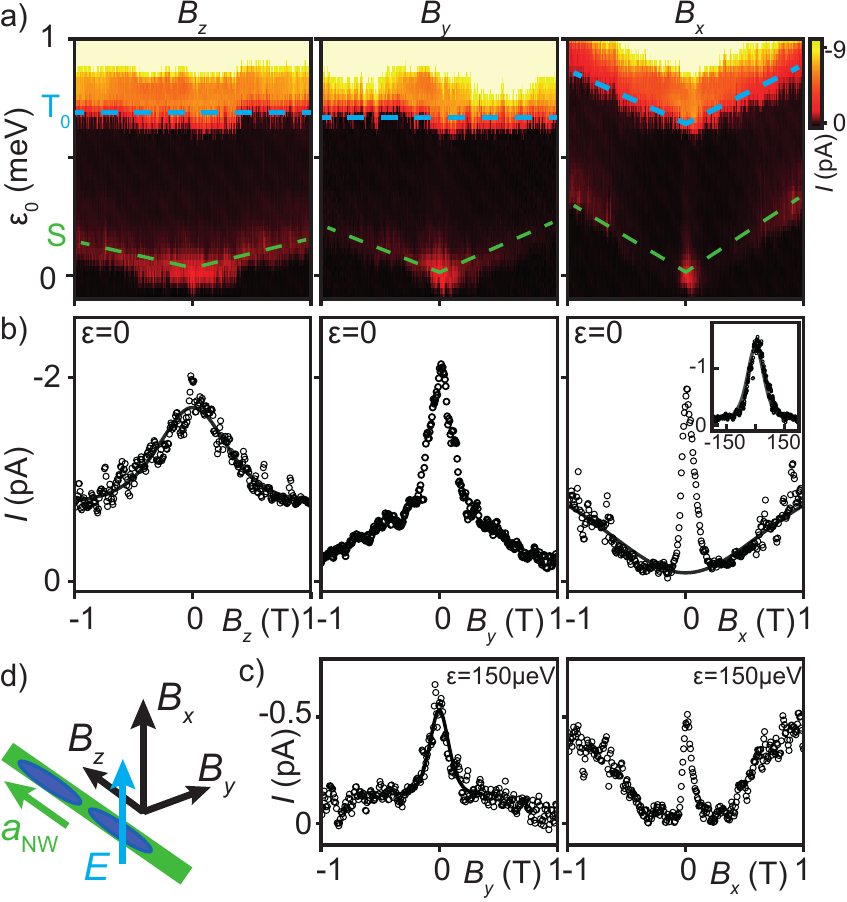}
\caption{(a) Magnetospectroscopy measurements for different magnetic field directions. $V_{g3}$ is swept from 486.0 to 489.0~mV while keeping $V_{g5} = 360.0$~mV fixed and sweeping $B$ from 1 to -1~T. (b) and (c) line cuts from a) at fixed $\epsilon$ (open circles) alongside fitted curves (solid lines). The inset shows a high-resolution scan ofthe central $I(B_x)$-peak, $B_x$ is here plotted in mT. (d) Schematic depiction of the magnetospectroscopy directions.}\label{Fig3}
\end{figure}

We start our discussion with $\vec{B}$ parallel to the nanowire axis. The zero-detuning line cut [left panel in Fig.~\ref{Fig3}(b)] has a maximum at $B=0$ and decreases for increasing magnetic field. Similar $I_\text{leak}(B)$ curves with peak widths of several hundred millitesla have been reported in other systems \citep{Lai2011,Yamahata2012,Li2015a} and were explained by spin-flip cotunneling \citep{Coish2011}. We can fit the peaks to
\begin{equation}\label{eq:cot}
I_{\text{co}}(B) = I_\text{res} + \frac{4 e c g^* \mu_{\text{B}} B}{3\sinh\frac{g^* \mu_{\text{B}} B}{k_{\text{B}} T}}\,,
\end{equation}
with $c = \frac{h}{\pi}[\{\Gamma_{\text{r}}/(\Delta-\epsilon)\}^2 + \{\Gamma_{\text{l}}/(\Delta + \epsilon -2 U_{\text{M}} - 2eV_{\text{SD}})\}^2]$, where $I_\text{res}$ is the residual leakage current, $e$ the electron charge, $g^*$ the effective $g$~factor, $h$ Planck's constant, $\Gamma_{\text{l,(r)}}$ the tunnel coupling with the left (right) reservoir, $\Delta$ the depth of the two-hole level, and $U_{\text{M}}$ the mutual charging energy \citep{Coish2011}. The fit in the left panel of Fig.~\ref{Fig3}b gives $g^*_{z} = 0.4\pm0.1$ with a hole temperature $T = 40$~mK. The finite residual current of $I_\text{res}=0.8$~pA at high magnetic fields indicates a second leakage process efficient at finite magnetic field, \eg spin-orbit interaction \citep{Danon2009}. The obtained $g$~factors for a magnetic field applied parallel to the nanowire axis are consistent with our findings for a single quantum dot \citep{Brauns2015}, where $g^*_\text{z} = 0.2\pm0.2$. 

\paragraph{}
To summarize, for $\vec{B}$ pointing along the nanowire axis we can explain the peak of $I_\text{leak}$ at $B=0$ with spin-flip cotunneling and obtain $g^*_z = \num{0.4\pm0.1}$, and we find hints for additional leakage processes above $\pm0.8$~T.

\subsection{Magnetospectroscopy along $B_y$}
\label{sec:leak_perpE_perpNW}
We now let the magnetic field point perpendicular to both the nanowire and the electric field. The zero-detuning line cut suggests the superposition of at least two peaks: a narrow peak dominates up to $B \approx 0.2$~T, and is superimposed on a broader peak that is visible over the whole range from --1 to 1~T and cannot be explained without further investigation. At finite detuning [left panel in Fig.~\ref{Fig3}(c)] this broader peak is not observed, which permits us to perform a more precise fit of the central peak. By fitting to Eq.~(\ref{eq:cot}) we obtain $g^*_{\text{y}} = 1.2\pm0.2$ at $\epsilon = 150$~$\mu$eV. The $g$~factor along this magnetic field direction is significantly lower than in single quantum dot measurements \citep{Brauns2015}, where $g^*_{\text{y}} = 2.7\pm0.1$ was maximal. 
\paragraph{}
Because $g_y>g_z$, the spin-flip cotunneling induced leakage current is exponentially suppressed for magnetic fields above $B \approx 0.25$~T and we obtain $I_\text{res}$ at $\epsilon = 150$~$\mu$eV of $\sim -0.1$~pA. At zero detuning, the minimum observed current is $\sim -0.2$~pA, which might be overestimated because of the additional features at high magnetic fields.
\paragraph{}
In summary, for $B \,{\perp}\, E$, $\,{\perp}\, \text{NW}$ we observe a peak in the leakage current at $B = 0$, which we explain with spin-flip cotunneling and find $g^*_z \approx \num{1.2}$. The remaining leakage current is significantly lower than for $\vec{B} \parallel \vec{a}_\text{NW}$.

\subsection{Magnetospectroscopy along $B_x$}
\label{sec:leak_parE_perpNW}
The third high-symmetry direction is $B_x$, parallel to $E$ and perpendicular to the nanowire. Similar to the other two magnetic field directions we find a peak around $B = 0$ in the $I_{\text{leak}}(B)$-curve at $\epsilon = 0$ [right panel of Fig.~\ref{Fig3}(b)]. We fit Eq.~(\ref{eq:cot}) to a high-resolution magnetic field sweep at a ten times lower sweeping rate (5~mT/min instead of 50~mT/min) [inset of right panel in Fig.~\ref{Fig3}(b)], which results in $g^*_x = 3.9\pm0.1$. $g^*_x$ here is substantially higher than the $g^*_y$ values obtained for $B \perp E$, as opposed to our findings in single quantum dots \citep{Brauns2015}, where $g^*_y > g^*_x$. One possible explanation for this discrepancy is that it is very likely that we do not operate in the lowest-energy subband of the nanowire. Here, we estimate approximately 70 holes to reside in each quantum dot. The theoretical calculations for the $g$~factor anisotropy \citep{Maier2013} that we have confirmed experimentally take into account only the quasi-degenerate two lowest-lying subbands, and other theoretical work suggests that the heavy-hole light-hole mixing is very different for different subbands \citep{Csontos2008,Csontos2009}. The measured angle dependence of $g^*$ confirms that the measured $g^*$-factor anisotropy is not related to the crystal orientation but to the one-dimensional confinement in the nanowire and a finite electric field perpendicular to the nanowire axis.
\paragraph{}
The leakage current at magnetic fields $B > 0.1$~T, where spin-flip cotunneling is efficiently suppressed, exhibits a qualitatively different behavior than for $B_y$: Up to $B \approx \SI{0.3}{\tesla}$ the leakage current is minimal and increases again for $B > \SI{0.3}{\tesla}$. Up to $B = 1$~T the leakage current does not fully saturate, although the slope reduces at both $\epsilon = 0$ and $\epsilon = 150$~$\mu$eV when $B$ approaches 1~T, which hints at a saturation for $B > 1$~T. Such an increasing $I_{\text{leak}}(B)$ with saturation at higher $B$ is again an indication for spin-orbit induced leakage. Taking the line trace at $\epsilon=0$ in the right panel of Fig.~\ref{Fig3}(b), we find a minimal leakage current of $I_\text{min} = 0.1\pm0.1$~pA. The leakage current due to spin-orbit coupling between (2,0) triplet and (1,1) singlet states can be expressed as
\begin{equation}\label{eq:SOI}
I_{\text{SOI}} = I_{\text{max}}\left(1-\frac{8B_{\text{C}}^2}{9 (B^2 + B_{\text{C}}^2)}\right)\,,
\end{equation}
where $I_{\text{max}}$ is the maximum leakage current at high magnetic fields, and $B_{\text{C}}$ the width of the characteristic dip around $B=0$ \citep{Danon2009}. If we now assume the minimum leakage current $I_{\text{min}}$ to be exclusively due to a spin-orbit interaction we expect $I_{\text{max}} = 9 I_{\text{min}} = 1\pm1$~pA at high fields. This estimate is in reasonably good agreement with the value for $I_{\text{max}}$ we obtain by fitting the data excluding the central peak to Eq.~(\ref{eq:SOI}) [see red solid line in the right panel of Fig.~\ref{Fig3}(b)], where we find $I_{\text{max}} = -1.6\pm0.3$~pA and $B_\text{C} = 1.0\pm0.2$~T. Since at zero detuning $I_{\text{max}} = 4 e \Gamma_\text{rel}$, where $\Gamma_\text{rel}$ is the spin relaxation rate \citep{Danon2009}, we can calculate $\Gamma_\text{rel} = 2.5\pm0.5$~MHz. This is comparable with reports on measurements of heavy holes in intrinsic Si \citep{Li2015a}, where $\Gamma_\text{rel} = 3$~MHz, and electrons in InAs \citep{Nadj-Perge2010}, where $\Gamma_\text{rel}$ ranges from 0 to 5.7~MHz. We note that $B_{\text{C}}$ is of the order of $\sim 1$~T. \citet{Danon2009} neglect the Coulomb interaction effects of $B$, which are of relevance at such field strengths (see also the Supplemental Material S1). Therefore also other spin-orbit or Coulomb related effects can provide significant leakage paths \citep{Khaetskii2000,Golovach2004a,Badescu2005,Flindt2006,Trif2007}.
\paragraph{}
To sum up, the spin-flip cotunneling peak in $I_\text{leak}(B,\epsilon\!=\!0)$ can be efficiently quenched by a magnetic field due to a very high effective $g$~factor of $g^*_x \approx 3.9$. $g^*_x$ is significantly higher than $g^*_y$, in contrast to our findings in single quantum dots, where we see the opposite behavior. At higher $B$ we notice an increasing $I_\text{neg}$, which we assign to spin-orbit coupling induced mixing of the spin states.
\paragraph{}
Let us now briefly compare our findings with existing literature and discuss the implications for spin-orbit qubits. \citet{Pribiag2013} measured the leakage current at three different angles in plane of the sample and found an almost isotropic dependence of the SOI-induced leakage current. To our knowledge there are no reports of an angle dependence in the plane perpendicular to the nanowire. Spin-flip cotunneling limited leakage current as found in our device is exponentially suppressed by a Zeeman splitting of the spin states at finite $\vec{B}$, \ie the remaining leakage current at a given $\vec{B}$ depends on the effective $g$~factor. Since we find $g^*$ to be highly anisotropic with respect to both $\vec{E}$ and $\vec{a}_{\text{NW}}$, the leakage current can be minimized by pointing $\vec{B}$ along $\vec{E}$. Also the leakage current caused by SOI is anisotropic and depends on the wave function overlap between the two dots \citep{Danon2009}. Therefore it is possible to tune the double quantum dot so that the SOI leakage current dip around $B=0$ is wider than the leakage current peak around $B=0$ caused by spin-flip cotunneling, which is the case here for $\vec{B} \parallel \vec{E}$. Previously measured spin relaxation times of several hundred $\mu$s obtained with $\vec{B}$ along the nanowire \citep{Hu2012} might be extended by an order of magnitude when measured parallel to  $\vec{E}$ according to the data presented here.

\section{Conclusion}

In conclusion, we have successfully formed a gate-defined double quantum dot in a Ge-Si core-shell nanowire. We have observed shell filling and a Pauli spin blockade, and have been able to explain the observed leakage current by a combination of spin-flip cotunneling at low magnetic fields and SOI-induced coupling between singlet and triplet states at higher fields.
\paragraph{}
With these results we show that by wisely choosing the magnitude and direction of a magnetic field applied to a Pauli spin-blocked double quantum dot, one can achieve longer spin lifetimes.

\begin{acknowledgments}
We thank Stevan Nadj-Perge, Sergey Amitonov, and Paul-Christiaan Spruijtenburg for fruitful discussions. We acknowledge technical support by Hans Mertens. F.A.Z. acknowledges financial support through the EC Seventh Framework Programme (FP7-ICT) initiative under Project SiAM No. 610637, and from the Foundation for Fundamental Research on Matter (FOM), which is part of the Netherlands Organization for Scientific Research (NWO). E.P.A.M.B. acknowledges financial support through the EC Seventh Framework Programme (FP7-ICT) initiative under Project SiSpin No. 323841.
\end{acknowledgments}

\end{document}



\title{Anisotropic Pauli spin blockade in hole quantum dots\newline Supplementary Material}


\author{Matthias Brauns}
\email[Corresponding author, e-mail: ]{m.brauns@utwente.nl}
\author{Joost Ridderbos}
\affiliation{NanoElectronics Group, MESA+ Institute for Nanotechnology, University of Twente, P.O. Box 217, 7500 AE Enschede, The
Netherlands}
\author{Ang Li}
\affiliation{Department of Applied Physics, Eindhoven University of Technology, Postbox 513, 5600 MB Eindhoven, The Netherlands}
\author{Erik P. A. M. Bakkers}
\affiliation{Department of Applied Physics, Eindhoven University of Technology, Postbox 513, 5600 MB Eindhoven, The Netherlands}
\affiliation{QuTech and Kavli Institute of Nanoscience, Delft University of Technology, 2600 GA Delft, The Netherlands}
\author{Wilfred G. van der Wiel}
\affiliation{NanoElectronics Group, MESA+ Institute for Nanotechnology, University of Twente, P.O. Box 217, 7500 AE Enschede, The
Netherlands}
\author{Floris A. Zwanenburg}
\affiliation{NanoElectronics Group, MESA+ Institute for Nanotechnology, University of Twente, P.O. Box 217, 7500 AE Enschede, The
Netherlands}


\date{\today}


\pacs{}

\maketitle


\section{S1 - Anisotropic Coulomb interaction}

We start our discussion with the magnetospectroscopy measurements for $B_z$ and $B_y$ (left and middle column in Fig.~3a of the main text). In supplementary Fig.~1 we plot $I(\epsilon_0)$ line cuts at $B=0$~T and $B=1$~T. The \Tn-onset only changes minimally between 0 and 1~T, whereas the S-onset shifts to positive $\epsilon_0$-values by $\Delta\text{S}= 100\pm10$~$\mu$eV for $B_z$ and by $\Delta\text{S}= 200\pm10$~$\mu$eV for $B_y$. As stated in the main text this behaviour has also been observed in other experiments \citep{Kyriakidis2001,Fujisawa2003,Johnson2005}. It can be explained by orbital effects of the magnetic field \citep{Wagner1992,VanderWiel1998,Kyriakidis2001}, where $B$ leads to a shrinking of the wave function which increases the Coulomb energy cost of singlet states as compared to triplet states and therefore leads to a shift of the S-onset to higher energies.

\begin{figure}
  \includegraphics{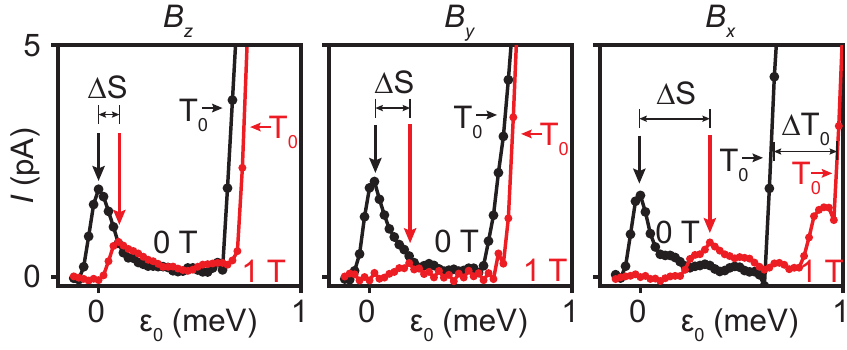}
\caption{$I(\epsilon_0)$ line cuts from Fig.3 in the main text at $B=0$~T (black curve) and $B=1$~T (red) for all three magnetic field directions. $\Delta \text{S}$ denotes the shift of the S-onset, $\Delta \text{T}_0$ the shift of the T$_0$-onset.}\label{SFig1}
\end{figure}

\paragraph{}
A peculiarity of the measurements for $B_x$ is the \Tn-onset, which, in contrast to the measurements along $B_y$ and $B_z$,  shifts to higher $\epsilon_0$ for increasing $B$. Between $B=0$ and $B=\SI{1}{\tesla}$ this shift amounts to $\Delta T_0 = 280\pm10$~$\mu$eV, whereas the shift of the S-onset sums up to $\Delta S = 330\pm10$~$\mu$eV. For $B_z$ and $B_y$, we only find $\Delta S = 100\pm10$~$\mu$eV and $\Delta S = 200\pm10$~$\mu$eV, respectively. This enhanced $\Delta S$ along with the pronounced $\Delta T_0$ for relatively small magnetic fields $B \le 1$~T indicate a change in the Coulomb interaction between the two holes as compared to the measurements for $B_y$ and $B_z$ \citep{Kyriakidis2001,Kouwenhoven2001}. Theoretical calculations suggest an anisotropic character of hole-hole interactions with respect to the electric-field axis in Ge-Si core-shell nanowires \citep{Maier2014b}, but further studies are necessary for a definite conclusion on the cause for the observed anisotropic $B$-dependence of the two-hole spin states. 

\section{S2 - Tunnel couplings}

\begin{figure}
  \includegraphics{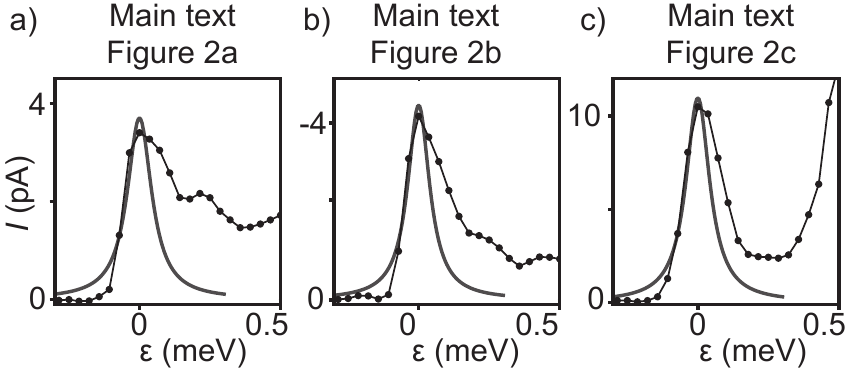}
\caption{Zooms of the $I(\epsilon)$ line cuts for the unblocked bias direction in the lower panels of Fig.~2 in the main text.}\label{SFig2}
\end{figure}

From the $I(\epsilon)$-curves in the non-blocked bias direction (lower panels of Fig.~2 in the main text), we can extract the tunnel coupling $\Gamma$ to the reservoirs and the interdot tunnel coupling $t$ by fitting to the expression 
\begin{equation}\label{eq}
I(\epsilon) = (\Gamma \abs{t}^2)/((\epsilon/h)^2 + \Gamma^2/4 + 3\abs{t}^2)\,,
\end{equation}
where $h$ is the Planck constant \citep{Wiel2002}. We fit the $I(\epsilon)$-curves to Equation~\ref{eq} only for $\epsilon \le 0$, because for $\epsilon > 0$ inelastic processes that are not taken into account lead to an increased current, as can be inferred from the asymmetry of $I(\epsilon)$ around $\epsilon=0$ in supplementary Fig~\ref{SFig2} where we plot the fits alongside the data (the same approach has been used e.~g. by \citet{Li2015a}). The extracted values for $\Gamma$ and $t$ are:
\begin{itemize}
\item[a)] $\Gamma = 100\pm10$~$\mu$eV, $t = 1.6\pm0.2$~$\mu$eV
\item[b)] $\Gamma = 90\pm10$~$\mu$eV, $t = 1.6\pm0.2$~$\mu$eV
\item[c)] $\Gamma = 100\pm10$~$\mu$eV, $t = 2.7\pm0.2$~$\mu$eV
\end{itemize}
These values suggest that by changing the barrier voltages between the left and the right panel the interdot tunnel coupling $t$ increases significantly from 1.6~$\mu$eV to 2.7~$\mu$eV. This increase reflects an increased wavefunction overlap between states in the left and the right dot, which can be explained by the finite cross capacitance between g2/g6 and the quantum dots which may affect the shape of the wave functions. In all three cases $t$ is more than an order of magnitude smaller than $\Gamma$ and thus the dominant factor for the measured current.

\paragraph{}
Let us now take a more careful look at how the tunnel couplings relate to the leakage current in Pauli spin blockade. Although $t$, which is the tunnel coupling limiting the current through the double quantum dot,  is significantly higher in the right panel than in the left panel, $I_{\text{leak}}$ even \emph{decreases} (see Fig.~2 in the main text). This means that an increased overlap between (1,1) and (2,0) states does not play a major role in the process that determines the leakage current.